\documentclass[aps,prb,preprint,showpacs, showkeys]{revtex4}
\usepackage{graphics}
\usepackage{graphicx}
\usepackage{amsmath}
\usepackage{amssymb}
\usepackage{bm}
\usepackage{dcolumn}
\usepackage{epstopdf}
\begin{document}  

   \title[]{Optical Properties of Isolated and Supported Metal Nanoparticles}   
   \author{Cecilia Noguez}
   \email{cecilia@fisica.unam.mx} 
   \affiliation{Instituto de F\'{\i}sica, Universidad Nacional Aut\'onoma de M\'exico, Apartado Postal   20-364, Distrito 
   Federal 01000,  M\'exico}

   \date{Accepted in Optical Materials, Nov. 22 (2004)}  

 \begin{abstract}
A review of the main phenomena related with the linear optical properties of isolated and supported metal nanoparticles is presented. The extinction, absorption and scattering efficiencies are calculated using the Mie theory and the Discrete Dipole Approximation. The origin of the optical spectra is discussed in terms of the size, shape and environment for each nanoparticle. The main optical features of each nanoparticle are identified, showing the tremendous potentiality of optical spectroscopy as a tool of characterization.
 \end{abstract}

  \pacs{78.67.-n  }
  \keywords{Optical Properties, Metal Nanoparticles, Mie Theory, Discrete Dipole Approximation, Colloidal Nanoparticles, Supported Nanoparticles}

 \maketitle 

\section{Introduction}
\label{intro}

Shape, size and ambient conditions are crucial parameters to understand the physics and chemistry phenomena of matter at the nanometer scale. The promising technological applications of Nanoscience depend on our capacity to control these parameters. In particular, the correct knowledge of the optical properties of nanoestructures can be applied to fabricate new optoelectronic devices~\cite{1}, as well as a tool for characterization~\cite{2}. Optical techniques, like Raman, Differential and Anisotropy Reflectance Spectroscopy, Light Absorption Spectroscopy, Surface Enhanced Raman Scattering (SERS), etc., can be powerful tools to characterize nanostructures because of their non-destructive and real-time character together with their in situ potentiality. Furthermore, optical spectroscopies provide statistical properties of the whole sample. These attributes have allowed us to control the growth of superlattices~\cite{3}, and with a proper implementation of them, it might be possible to control also the growth of NPs (NPs), correcting their shape and size during the process. Notice that optical spectroscopies can be used also as complementary tools of the structural characterization techniques like Atomic Force Microscopy (AFM), Scanning Tunneling Microscopy (STM), Transmission Electron Microscopy (TEM), etc., which provide the image of a small piece of the sample, giving information about local properties and characterizing a few NPs at a time. Additionally, these structural techniques have other limitations because in most cases the growth and characterization are made in different ambient, which is a serious problem since the properties of NPs are ambient dependent. In some of these techniques the sample is literally touched during characterization and sometimes this might substantially modify the properties of a nanoparticle. Furthermore, the growth and characterization of NPs are usually made at different times, and this might become an additional uncontrollable variable. In conclusion, the actual correct determination of the size and shape parameters of a given nanoparticle is still controversial because a more complete experimental determination is needed, together with a corresponding clear physical interpretation.
In the present work, we study the optical properties of metallic nanometer-sized particles with different shapes and sizes. We calculate and discuss spectra for the extinction, absorption and scattering efficiencies. In particular, we show results for silver NPs. First we study the case of isolated or suspended NPs, and then we consider the case when the nanoparticle is lying over a substrate.

\section{Size effects}

When a particle is excited by an electromagnetic (EM) field, its electrons start to oscillate with the same frequency as the one of the incident EM. The excited charges can transform energy from the incident EM wave into, for example, thermal energy in a so-called absorption process. However, the charges can be also accelerated such that they can radiate energy in any direction in a so-called scattering process. The sum of both effects is called the light extinction. Then, the extinction cross section is defined as the total EM power crossing the surface of the particle as a function of the irradiance of the incident light~\cite{4}. 

In this work, we consider NPs, which are large enough to employ the classical EM theory. However, they are still enough small to observe the dependence of its optical properties as a function of it size and shape. This means that the inhomogeneities of the particle is small compared to the wavelength of the incident field, such that, each point of the nanoparticle can be described in terms of its macroscopic behavior: the optical properties of the nanoparticle depend on the frequency only. Here, we restrict ourselves to the elastic or coherent case, where the frequency of the absorbed and scattered light is the same as the frequency of the incident light.

To understand the range of the different process as a function of the size of the particle, we first study the absorption and scattering phenomena of a spherical particle using the Mie theory~\cite{4}.  To compare the scattering or absorption process we employ the scattering, absorption and extinction efficiencies, which are dimensionless functions. In all of our calculations we consider silver NPs describe by dielectric function measured by Johnson and Cristy~\cite{5}. 

In Fig. 1 we show the (a) absorption and (b) scattering efficiencies as a function of the wavelength of the incident light, calculated using Mie theory for spherical silver NPs of different radii a in vacuum. In the spectra we can observe that at about 320nm all the efficiencies have a local minimum that corresponds to the wavelength at which the dielectric function of silver, both real and imaginary parts, almost vanish. Therefore, this feature of the spectra is inherent to the material properties and, as we observe below, it is independent of the particle geometry. Below 320nm, the absorption of light is mainly due to the intra-band electronic transitions of silver, therefore, this feature of the spectra should be also quite independent of the shape and size of the particles, as it is actually corroborated in all the graphs shown below corresponding to a silver particle. We observe that for particles with a²10nm the absorption process dominates the optical spectra. For these NPs, the absorption efficiencies show a single peak always located at ?=355nm, and their intensity increases linearly with the radii of the sphere. Also for a²10nm, we observe that the scattering effects are not present, however, as the size of the sphere increases light scattering becomes important rapidly.   
\begin{figure}[h]
\caption{(a) Absorption and (b) scattering efficiencies for spherical nanoparticles of different radii.}
\end{figure}
Let us analyze three different absorption phenomena related with the size of the particles. This will help us to understand the line-shape of the spectra. The first one is due to plasma resonances of the metal, the second is due to the relation between the electron mean free path and the finite size of the particle, and the third is due to electron damping by radiation effects.

\subsection{Plasma resonances effects}
 When the size of a homogeneous particle is much smaller than the wavelength of the incident light, $a\ll \lambda$ , the electronic cloud of the NP is displaced with respect to the ions of the system.  In the case of a homogeneous sphere, this displacement corresponds to a dipole charge distribution. The extinction and absorption cross sections in the dipolar approximation are given by~\cite{5}:
\begin{eqnarray}
\label{coeficientes}
C_{\rm ext} & = & \frac{8\pi n_{\rm med}}{\lambda}{\rm Im}\left[\alpha_d(\lambda)\right] \, , \\ \nonumber
C_{\rm abs} & = & \frac{16\pi a^2}{3}\left(\frac{ 2 \pi n_{\rm med}}{\lambda}\right)^4 |\alpha_d(\lambda)|^2 \, ,
\end{eqnarray}
where
\begin{equation}
\label{alfa}
\alpha_d(\lambda) = \frac{\epsilon_{\rm esf} - \epsilon_{\rm med}}{\epsilon_{\rm esf} + 2 \epsilon_{\rm med}}, \nonumber
\end{equation}
$\alpha_d(\lambda)$ is the dipolar polarizability of the sphere, $\epsilon_{\rm esf}$ is the sphere dielectric function, and $\epsilon_{\rm med}$ is the dielectric function of the surrounding media. In this case, one can say that the sphere has one polarization mode whose wavelength is given by the pole of the polarizability in Eq. (1), i. e., when $\epsilon_{\rm esf} + 2 \epsilon_{\rm med} = 0 $.
We obtain that this mode is independent of the size sphere, corroborating the results shown in Fig. 1(a) for $ a\leqslant10$~nm. In conclusion, we find that the dipolar description of the spectra is valid only for spheres with radii smaller than 10~nm. 

When the size of the sphere increases the displacement of the electronic cloud is not more homogeneous, such that, high-multipolar charge distributions can be induced. In this case, the absorption spectrum shows more than one polarization modes and usually the spectrum is broadening and becomes asymmetric.  Also, when the size of the sphere increases scattering effects are present which also make broader the spectra and move the dipolar mode to larger wavelengths. In Fig. 2, we show the absorption efficiency for spherical NPs with 7 and 40~nm of radii, calculated using Mie theory. We also show the quadrupolar mode contribution to the spectra. We observe that for particles with $a \leqslant 30$~nm the contribution of the quadrupolar mode is not present. When the size of the sphere increases, the quadrupolar mode appears at smaller wavelength than the dipolar mode, giving rise to a shoulder in the spectrum at $\lambda=345$~nm, for the sphere with $a=40$~nm, as shown in Fig. 3(b). The quadrupolar mode does not depend also on the size of the particle; so, the wavelength of this mode is always the same.
\begin{figure}[h]
\caption{Absorption efficiencies for spherical nanoparticles of different radii.}
\end{figure}

\subsection{Surface dispersion effects}
 The dielectric function of a metal can be described as
 \begin{equation}
\label{drude}
\epsilon(\omega) = \epsilon_{\rm intra} + \left( 1 - \frac{\omega_p^2}{\omega^2 + i \omega \gamma_M} \right),
\end{equation}
where $ \epsilon_{\rm intra}$ is the contribution from the intraband transitions, and the term in parenthesis is the contribution from the conduction electrons given by the Drude formula~\cite{7}. Here, 
$\omega_p$ is the plasma frequency, and $\gamma_M$ is the macroscopic damping constant due to the dispersion of the electrons by the ions of the system. The latter constant is directly related with the mean free path of the conduction electrons. When the dimension of the NP is smaller than the mean free path, the conduction electrons are also dispersed by the surface of the NP. To take into account the finite size of the NP, we can add, to the damping constant, a term that depends on the size of the particle like $\gamma=\gamma_M+\gamma(a)$. 
Then, the dielectric function of the NP is now described as
\begin{equation}
\label{drude2}
\epsilon(\omega, a) = \epsilon_{\rm intra} + \left( 1 - \frac{\omega_p^2}{\omega^2 + i \omega (\gamma_M+ \gamma)} \right).
\end{equation}
For spherical NPs, the damping term due to surface dispersion~\cite{8} is: $\gamma(a)= 3 v_F / 4 a$ , where $v_F$  is the Fermi velocity. As smaller is the sphere, the surface dispersion effects are more important. The surface dispersion does not change the wavelength of the proper modes but it only affects its intensity. The latter is observed in the adsorption efficiency for a small sphere ($a \leqslant 10$~nm), where the dipolar mode stays at the same wavelength but the peak is less intense and is symmetrically broadening. In Fig. 3 we show the extinction efficiency with and without surface damping effects for spheres of $a=5$~nm and $a=50$~nm. For $a \geqslant 50$~nm, the surface dispersion effects can be neglected.
\begin{figure}[h]
\caption{Extinction efficiencies with and without surface dispersion effects for spherical nanoparticles of different radii.}
\end{figure}

\subsection{Radiation damping effects}
 When the size of the particle is large enough, the electrons can be accelerated in the presence of the incident light, and then, they radiate energy in all directions. Because of this secondary radiation, the electrons lose energy experimenting a damping effect. Let us suppose that at the center of the sphere, its total dipole moment is given by $\vec{P} = \alpha \left[\vec{E}_{\rm inc} + \vec{E}_{\rm rad} \right]$, where $\vec{E}_{\rm inc}$ is the incident EM field, and $\vec{E}_{\rm rad}$ is the radiated EM field. Considering a solid homogeneous sphere, one can suppose that each volume element contributes to the total dipole moment like $d\vec{p}(\vec{r}) = \vec{P} dV$, and that each one of these elements produces a depolarization field at the center of the sphere, which is given by $d \vec{E}_{\rm rad}$. We solve $\vec{P}$ in a self-consistent way, and find~\cite{9} that:
 \begin{equation}
\label{rad}
\vec{E}_{\rm rad} = \left[-\frac{4 \pi}{3} + k^2 \frac{4 \pi}{3} a^2 + ik^3 \frac{2}{3} \frac{4 \pi}{3} a^3 \right] \vec{P} \, ,
\end{equation}
with $k= 2\pi / \lambda$. The first term in the parenthesis of Eq. (4) is the usual factor due to polarization in the quasi-static limit. The second term comes from the depolarization field that depends on the ratio between the radius of the sphere and the wavelength of the incident light, $x = a/\lambda$. The third term is a damping effect due to the radiation of the dipole at the center of the sphere, and also depends on $x$.  The depolarization field shifts the position of the mode to the red (larger wavelengths), while the radiation damping reduces the intensity, makes broader and asymmetric the spectrum.   

In Fig. 4 we show the extinction efficiency calculated using Mie theory, and using the dipolar approximation from Eq. (1), and including the surface dispersion and radiation damping effects from Eqs. (3) and (4). We observe for a sphere with $a=20$~nm that the radiation damping shift the wavelength of the dipole mode, such that, the agreement with Mie theory is remarkable. For a sphere with $a=40$~nm the radiation effects are more evident because the dipolar mode is red-shifted, made wider and asymmetric. The shoulder below $\lambda=350$~nm for the sphere with $a=40$~nm, corresponds to the quadrupolar mode, as shown in Fig. 4(b).
\begin{figure}[h]
\caption{Extinction efficiencies calculated using Mie theory, and the dipolar approximation with and without radiation damping corrections, for spherical nanoparticles of different radii.}
\end{figure}

\section{Shape effects}
Let us study the optical response of NPs with different shapes. Now, we need to use numerical methods, since the Mie theory is exact only for spherical particles~\cite{4}. Because of the complexity of the systems being studied, efficient computational methods capable of treating large size materials are essential. In this work we employ the Discrete Dipole Approximation (DDA), which is a computational procedure suitable for studying scattering and absorption of EM radiation by particles with sizes of the order or less of the wavelength of the incident light. DDA has been applied to a broad range of problems~\cite{10}, including metal NPs~\cite{11} and their aggregates. The DDA was first introduced by Purcell and Pennypacker~\cite{12}, and has been subjected to several improvements, in particular those made by Draine, and collaborators~\cite{13}. Below, we briefly describe the main characteristics of DDA and its numerical implementation: the DDSCAT code. For a full description of DDA and DDSCAT, the reader may consult Refs. [12-14]. The main idea behind DDA is to approximate an object, in our case the nanoparticle, by a large enough array of polarizable point dipoles. Once the location and polarizability of each dipole are specified, the calculation of the scattering and absorption efficiencies by the dipole array can be performed, depending only on the accuracy of the mathematical algorithms and the capabilities of the computational hardware.

Let us assume an array of $N$ polarizable point dipoles located at $\{ \vec{r}_j\}, \,\, j=1,2,\dots, N$; each one characterized by a polarizability $\alpha_j$. The system is excited by a monochromatic incident plane wave $\vec{E}_{\mathrm{inc}}(\vec{r},t) = \vec{E}_{\mathrm{0}}e^{i\vec{k} \cdot \vec{r}-i\omega t}$, where $\vec{r}$ is the position vector, $t$ is time, $\omega $ is the angular frequency of the incident light. Each dipole of the system is subjected to an electric field that can be split in two contributions: (i) the incident radiation field, plus (ii) the field radiated by all the other induced dipoles. The sum of both fields is the so called local field at each dipole and is given by
\begin{equation}
\vec{E}_{i,\mathrm{loc}}=\vec{E}_{i,\mathrm{inc}}+\vec{E}_{i,\mathrm{dip}}=\vec{E}_{\mathrm{0}}e^{i\vec{k} \cdot \vec{r}_i}-\sum_{i\neq j}\mathbf{A}_{ij}\cdot \vec{P}_{j},
\end{equation}
where $\vec{P}_{i}$ is the dipole moment of the $i$th-element, and $\mathbf{A}_{ij}$ with $i \neq j$ is an interaction matrix. Once we solve the $3N$-coupled complex linear equations given by relation 
$\vec{P}_{i}=\alpha_{i}\cdot \vec{E}_{i,\mathrm{loc}}$, then, we can then find the extinction and absorption cross sections for a target in terms of the dipole moments.

Here, we present results for silver NPs with different geometries. We calculate the polarizability using the Clausius-Mossotti relation~\cite{7}, and the dielectric function as measured on bulk silver by Johnson and Christy~\cite{5}. In Fig. 5, we present the absorption efficiency of silver cubes with sides smaller than 10 nm, as well as for large cubes with sides of 80 and 160~nm. In all cases, the spectra show a rich structure of peaks, contrary to the case of the sphere with a single peak. These peaks are associated to the resonances inherent to the cubic geometry~\cite{15}, because in this case the charges can arrange in many different ways, even in the quasi-static limit. At least six different modes are observed, being the dipolar charge distribution the one that contributes more to the spectra. For the small cube, the peaks are sharper than those for larger cubes. This is because of the radiation damping effects, which moves the spectrum to larger wavelengths, make broader the peaks and diminish the intensity. For large cubes, the peaks at lower wavelengths increase their intensity, since high-multipolar contributions become important. However, the main features of the spectra due to the geometry of the particle are still there, as one can observe in Fig.~5.
\begin{figure}[h]
\caption{Absorption efficiencies for cube nanoparticles of different size.}
\end{figure}

In Fig. 6(a), we show the absorption efficiencies for a prolate spheroid with major to minor axis ratio of 2 to 1, and where the incident EM field is polarized along and perpendicular to its major axis. The major axis is 6~nm of length. As expected for small particles, the main contribution to the efficiencies comes from the excited surface plasmons, one for each polarization. The position of these surface plasmons depends on the axis ratio. For each polarization, the peak is given by a dipole mode, where the one for an external field parallel to the major axis, is more intense than the one for a field perpendicular. This shows that optical techniques are also sensitive to light polarization. When the size of the spheroid increases, we observe the radiation damping effects, as well as important contributions from quadrupolar charge distributions~\cite{11}, like in the case for the sphere. 
\begin{figure}[h]
\caption{Absorption efficiencies for (a) a prolate silver spheroid, and for (b) a cylinder and a disc.}
\end{figure}

In Fig. 6(b), we show the absorption efficiencies for a cylinder and a disc, when the incident EM field is parallel to their symmetry axis. We observe that their optical response is similar to the one for the spheroid using polarized light. However, the spectra in for each particle show a main peak (like for the spheroid) plus other contributions. These contributions come from the different modes inherent to the geometry. For example, we have different charge distributions at the plates of the cylinder and disc, which give rise to the small structure around each one of the main peaks.  These allow us to distinguish among NPs with different geometries.

\section{Substrate effects}

The knowledge of the optical properties of a particle located above a substrate can be used also as a tool to interpret optical spectra for the characterization of supported NP systems. The optical spectra of metallic NPs deposited on an insulating substrate is also characterized by the presence of resonances. The location and broadening of these resonances depend on the morphological and physical properties of the system. For example, they depend on the properties of the substrate because the particle interacts with the charges induced on the substrate. The interaction between different supported particles is also important, specially in the case of a high concentration of particles. However, the study of a single supported particle can be also performed in the dilute regime. In previous works, the presence of the substrate was included by taking a dipolar interaction between the particle and its image. However, the calculation of the interaction between a particle of finite size and its image requires the inclusion of multipolar interactions. This is because the field produced by the image, rather than being homogeneous over the size of the particle, as required by the dipolar approximation, it is strongly inhomogeneous, specially when the particle is close to the substrate. Different authors have included these multipolar interactions considering particles of different shapes~\cite{16,17,18,19}. However, the numerical complexity of the problem restricts tremendously the number of multipolar interactions taken to describe the system. This situation restricts the study to only a few specific systems.

The inclusion of multipolar interactions between the particle and its image gives rise to resonances additional to the dipolar one, which is the one which characterizes an isolated particle. It has been also shown that the location and strength of the multipolar resonances depend strongly on the properties of the substrate and the geometry of the system~\cite{19}. For example, when a particle is in close contact with the substrate smooth spectra are obtained. On the contrary, when the particle is located a certain distance above the substrate, a well defined structure of the particle is located a certain distance above the substrate, a well defined structure of resonances is obtained. The structure of these resonances is more evident when the contrast in the dielectric response of the ambient and the substrate increases~\cite{19}.

A powerful theoretical procedure has been developed to calculate the optical response of a particle-substrate system using a spectral representation~\cite{19}. The main advantage of this representation is that the strength and localization of the resonances, when given in terms of the spectral variable, are independent of the dielectric properties of the particle, but depend only on its shape and the dielectric properties of the substrate. With this procedure one is also able to include a larger number of multipoles, allowing the treatment of substrates with a larger contrast in the dielectric constant and particles closer to the substrate. In this work we consider more asymmetric oblate particle than in previous treatments [16-18] and we also study the changes in differential-reflectance spectra for particles made of different materials. The details of the formalism is elsewhere~\cite{19}.

In Differential Reflectance (DR) spectroscopy one compares the reflectance of the substrate-film system with the reflectance of the clean substrate, that is,
\begin{equation}
\label{ }
\frac{\bigtriangleup R}{R} = \frac{R_{\rm substrate + film} - R_{\rm substrate}}{R_{\rm substrate}}.
\end{equation}
When one considers that the film is constituted by a dilute distribution of particles, all located at the same distance from the substrate, one obtains, for $p$-polarized light, the  following 
\begin{equation}
\label{ }
\frac{\bigtriangleup R_p}{R_p} = \frac{16 \omega f_2 a}{c} \cos\theta \left[ \frac{(\epsilon_s - \sin^2\theta) \tilde{\alpha}_{||} - \epsilon_s^2\sin^2\theta \tilde{\alpha}_{\bot}}{(1 -\epsilon_s) ( \sin^2\theta - \epsilon_s\cos^2 \theta)} \right],
\end{equation}
where $\theta (=50^\circ)$ is the angle of incidence, $f_2$  is the two-dimensional filling fraction of particles, $c$ is the speed of light, and $ \tilde{\alpha}_{||}$  and $ \tilde{\alpha}_{\bot}$   are the effective polarizabilities of the particle, parallel and perpendicular to the substrate. We consider here the case of particles of free-electrons metals whose dielectric function can be described by the Drude model. For potassium the Drude parameters are $\omega_p =3.8$~eV and  $(\gamma_M \omega_p)^{-1} =0.105$. We also consider silver particles, and in this case we use, in our calculations, the experimentally-determined dielectric function~\cite{5}.

In Fig. 7 we show the DR spectra corresponding to particles of potassium and silver lying over a substrate of sapphire. In two of the panels, $a/b=1$ (spheres) while in the other two $a/b=2.5$. The spheres are located at $d=1.0005b$. Our procedure based in the spectral representation allows us to consider a number of multipolar contribution as large as $L_{\rm max}=2000$, which is the number required to obtain multipolar convergence in the calculation of the spectral function. In the left panels of Fig. 7, we observe how the final shape of the spectrum comes about by the superposition of resonances with strengths with different signs. In the right panels of Fig. 7 we also observe important differences in the shape of the spectra due to the difference in the dielectric response of the material the particles are made of, even when in both cases the resonance structure is the same. A more complex structure is observed in the case of particles of Ag. In potassium spheres the spectrum does not reflect such a rich resonance structure, and this is mainly because the particle lies too close to the substrate. From these differences in the spectra it is difficult imagine how an invariant feature could be obtained from them. The spectral representation is that invariant. We can give a more detailed explanation of the spectra of Fig.7 in the following way. In the case of silver spheres the resonances are taller and sharper than for potassium. As a consequence the resonance structure is richer than in potassium where the broadening effects wash out the details of this resonance structure. The peak at low frequencies is thinner and higher than that observed at higher frequencies, the opposite of what happens in the case of potassium, which arises from the specific combination of parameters. At low frequencies the resonances of each mode are taller and sharper than at high frequencies.

\begin{figure}[h]
\caption{DR for particles embedded in air lying over a substrate of sapphire. Left panels correspond to spherical particles at a distance $d=1.0005b$ from the substrate, while right panels correspond to OS particles with $a/b=2.5$ at a distance $d=1.05b$. Upper panels correspond to particles of potassium and lower panels to particles of silver.}
\end{figure}

\section{Conclusions}
We have shown that optical spectroscopies can be useful to characterize nanoparticles of different size, shape and ambient conditions. We also show that the spectra is sensitive to light polarization. In most cases, we have clearly identified the main optical signature associate to different geometries. We have identified in the spectra the main surface plasmon resonance associated to a dipolar excitation, as well as other resonances due to high-multipolar excitations. A direct comparison of our results with most of the available experimental measurements of the optical properties of suspended nanoparticles would require an averaging procedure over a wide distribution of sizes and shapes.  But this averaging procedure might smooth out the main relevant features of the spectra associated to the size and shape of the nanoparticles. On the other hand, it would be very desirable to obtain optical spectra over samples with narrower distributions of sizes and shapes.

\acknowledgments We thank to A. L. Gonz\'alez, C. E. Rom\'an, I. O. Sosa and R. G. Barrera for their contributions at different stages of this work. We also acknowledge to G. P. Ortiz for his comments. This work is partly supported by DGAPA-UNAM Grant No. IN104201, and CONACyT Grant No 36651-E.


\begin{thebibliography}{99}

 \bibitem{1}  V. R. Almeida, et al., Nature 421 (2004), p. 1081.
 \bibitem{2}  A. M. Rao, et al., Science 275 (1997), p. 187.
 \bibitem{3}  See for example: A. R. Turner, et al., Phys. Rev. Lett. 74 (1995), p. 3213; C. Noguez, et al., Phys. Rev. Lett.  76 (1996), p. 4923; B. G. Frederick, et al., Phys. Rev. Lett. 80 1998), p. 4490; J. R. Power, et al., Phys. Rev. Lett. 80 (1998), p. 3133.
 \bibitem{4} C. F. Bohren and D. R. Huffman, Absorption and Scattering of Light by Small Particles, John Wiley \& Sons, New York, 1983.
 \bibitem{5} P. B. Johnson and R. W. Christy, Phys. Rev. B 6 (1972), p. 4370. 
 \bibitem{6} K.L. Kelly, E. Coronado, L.L. Zhao, G.C. Schatz, J. Phys. Chem. B 107 (2003), p. 668.
 \bibitem{7} Classical Electrodynamics, 3rd edition, by J. D. Jackson, John Wiley \& Sons, NewYork, 1998.
 \bibitem{8}  U. Kreibig, J. Phys. F: Metal Phys.  4 (1974), p. 999.
 \bibitem{9}  M. Meier and  A. Wokaun,  Opt. Lett. 8 (1983), p.  581.
 \bibitem{10} Light Scattering by Nonspherical Particles, Edited by M. I. Mishchenko, J. W. Hovenier, and L. D. Travis, Academic Press, San Diego, 2000.
 \bibitem{11} I. O. Sosa, C. Noguez, and R. G. Barrera, J. Phys. Chem. B 107 (2003), p. 6269.
 \bibitem{12} E.M. Purcell and C.R. Pennypacker, Astrophys. J. 186 (1973), p. 705.
 \bibitem{13} B. T. Draine, Astrophys. J. 333 (1998), p. 848; B. T. Draine and J.J. Goodman, ibid  405 (1993), p.  685; B. T. Draine and P.J. Flatau, J. Opt. Am. A 11 (1994), p. 1491.
 \bibitem{14} Program DDSCAT, by B. T. Draine and P.J. Flatau, University of California at San Diego. 
 \bibitem{15} R. Fuchs, Phys. Rev. B 11 (1975), p. 1732. 
 \bibitem{16}	R. Ruppin, Surf. Sci. 127, 108 (1983).
 \bibitem{17}	M. M. Wind, J Vlieger and D. Bedeaux, Physica 141A, 33 (1987); M. M. Wind, P. A. Bobbert, J Vlieger and D. Bedeaux, Physica 143A, 164 (1987).
 \bibitem{18}	P. A. Bobbert and J. Vlieger, Physica 137A, 209, 243 (1986).
 \bibitem{19}	C. E. Roman-Velazquez, C. Noguez and R. G. Barrera, Phys. Rev. B 61 (2000), p. 10427.
\end{thebibliography}
\end{document}